\documentclass[final,5p,times,twocolumn]{elsarticle}

\usepackage{float}
\usepackage{subfig}
\usepackage{lineno,hyperref}
\usepackage{cleveref}

\journal{Nuclear Instruments and Methods in Physics Research Section A}

\begin{document}

\begin{frontmatter}

\title{Plasma ramps caused by outflow in gas-filled capillaries}

\author[LNF]{F. Filippi\corref{mycorrespondingauthor}}
\cortext[correspondingauthor]{Corresponding author}
\ead{francesco.filippi@roma1.infn.it}

\author[LNF]{M. P. Anania}
\author[LNF]{A. Biagioni}
\author[LNF]{E. Brentegani}
\author[LNF]{E. Chiadroni}
\author[TorVergata]{A. Cianchi}
\author[LNF]{M. Ferrario}
\author[LNF]{A. Marocchino}
\author[HebrewUniversity]{A. Zigler}

\address[LNF]{Laboratori Nazionali di Frascati, INFN, Via E. Fermi, Frascati, Italia}
\address[TorVergata]{Dipartimento di Fisica, Universit\'a di Roma Tor Vergata, Via della Ricerca Scientifica 1, 00133 Roma, Italia}
\address[HebrewUniversity]{Hebrew University of Jerusalem, Jerusalem 91904, Israel}

\begin{abstract}
Plasma confinement inside capillaries has been developed in the past years for plasma-based acceleration to ensure a stable and repeatable plasma density distribution during the interaction with either particles or laser beams. In particular, gas-filled capillaries allow a stable and almost predictable plasma distribution along the interaction with the particles. However, the plasma ejected through the ends of the capillary interacts with the beam before the inner plasma, affecting the quality of the beam. In this article we report the measurements on the evolution of the plasma flow at the two ends of a 1 cm long, 1 mm diameter capillary filled with hydrogen. In particular, we measured the longitudinal density distribution and the expansion velocity of the plasma outside the capillary. This study will allow a better understanding of the beam-plasma interaction for future plasma-based experiments.
\end{abstract}


\end{frontmatter}


\section{Introduction}
Compared with other plasma sources, gas filled capillaries allow to control the properties of the plasma along the entire length of the capillary, letting to properly set up the correct plasma distribution suitable for plasma acceleration experiments \cite{Spence2000,Bobrova2001}. In these devices the gas inside the capillary is ionized by a discharge current of several hundreds of amperes. This allows for centimeter-long and relatively controllable plasma channel which have been used in several experimental campaigns \cite{Leemans2006,Gonsalves2011,VanTilborg2015,Pompili2017}. However, during the ionization process and afterwards the plasma density evolves modifying its value and its distribution from the beginning of the discharge till the complete recombination \cite{FilippiIPAC2017}. Part of the plasma flows out the capillary from its ends decreasing the plasma density inside and producing a plasma density ramp which extends the length of the plasma beyond the capillary dimension. The effective plasma density profile encountered by the laser beam or the electron beam is different from the one generated only inside the capillary. Previous studies showed the effects of a low density plasma profile flown out the capillary on the electron bunch dynamics \cite{Marocchino2017}. Moreover, due to the different light propagation properties, outer plasma density may modify the laser pulse before the interaction with the inner plasma \cite{Kruer1988}. It is straightforward that the plasma outside the capillary is not negligible for the quality of the plasma-accelerated particles and its effect must be taken into account.
\begin{figure}
	\begin{center}
		\includegraphics[width=0.8 \columnwidth, trim=0 0 0 0]{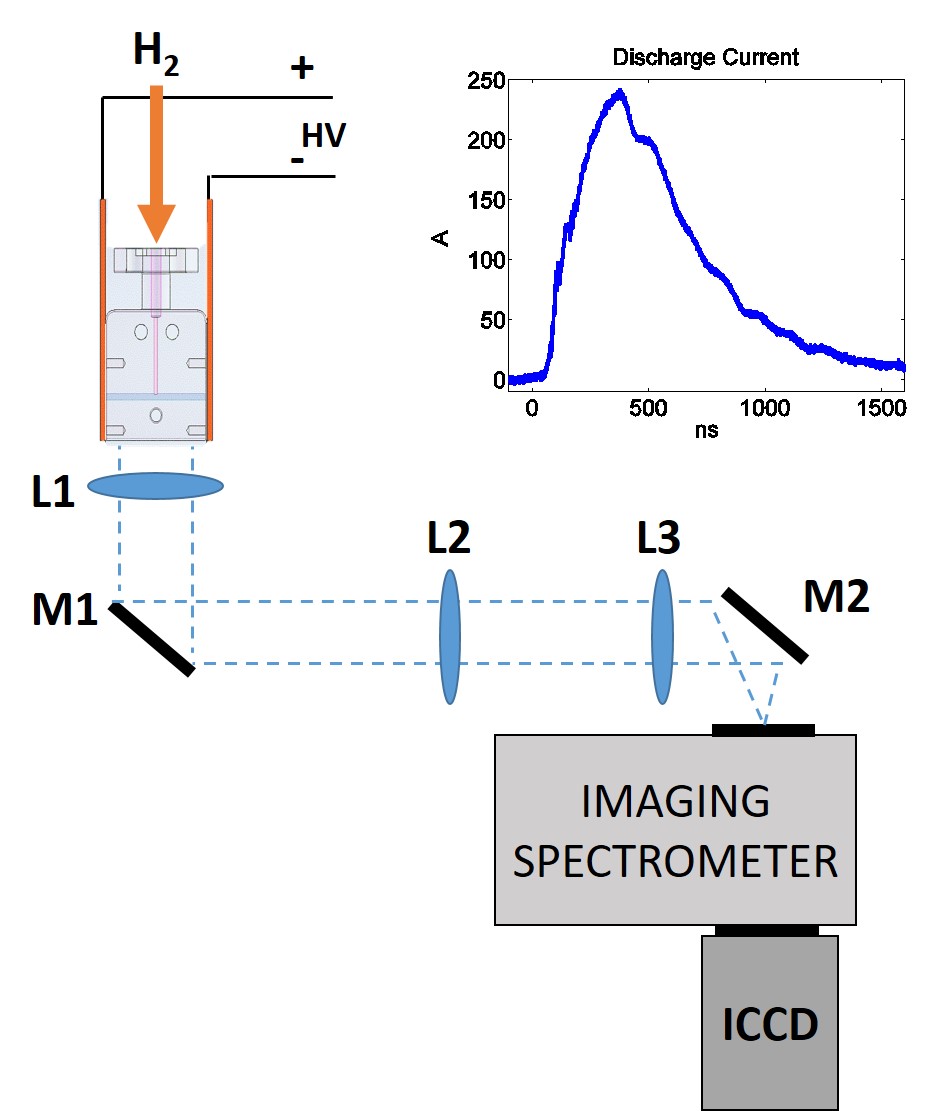}
		\caption{\label{CapillarySchematic}Experimental layout of the plasma density measurements.}
	\end{center}
\end{figure}

In this article, we present the measurements of the evolution of the plasma density in a 3D-printed cylindrical capillary of 1 cm length and 1 mm of diameter filled from a single central inlet with hydrogen and ionized by a discharge current of 230 A of peak value, as represented in Fig. \ref{CapillarySchematic}. This experimental setup has been studied for plasma based experiment at SPARC\_LAB and has been used in recent measurements \cite{PompiliNIMA2017}. We measured the electron plasma density by measuring the Stark broadening of the Balmer beta line (486.1 nm) collected and imaged onto an imaging spectrometer with a temporal resolution of 100 ns and spatial resolution of 133 $\mu$m \cite{FilippiJinst2016}. By measuring the full width at half maximum (FWHM) of the Balmer beta line it is possible to retrieve the electron plasma density localized around the emitter \cite{Gigosos2003}. During our measurements we observed the evolution of the inner plasma density as well as the evolution of the outer plasma density.

\section{Gas-filled capillaries for plasma accelerators}
\begin{figure}
	\begin{center}
		\includegraphics[width=1 \columnwidth, trim=0 0 0 0]{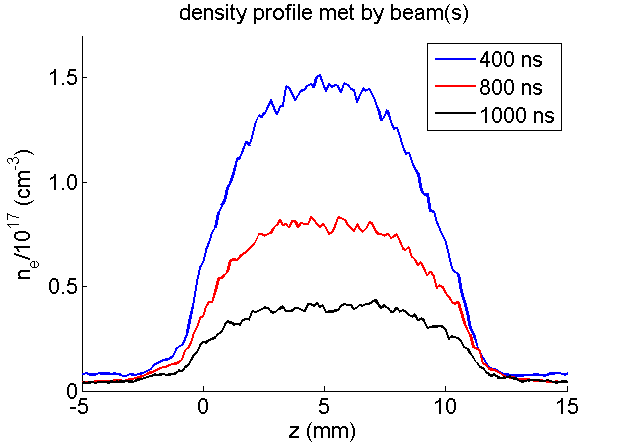}
		\caption{\label{DensityProfile}Density profiles acquired 400, 800 and 1000 ns after triggering of the discharge. The capillary extend from 0 to 10 mm.}
	\end{center}
\end{figure}
\begin{figure*}[!h]
	\centering
	\includegraphics[width=0.26 \paperwidth]{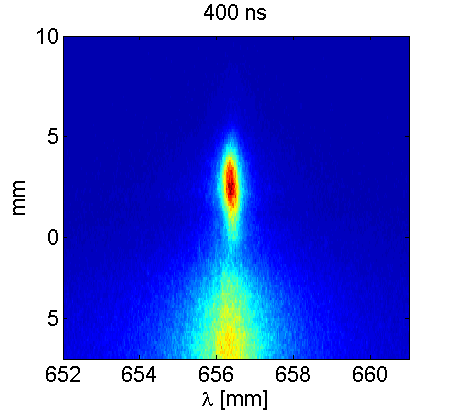}
	\includegraphics[width=0.26 \paperwidth]{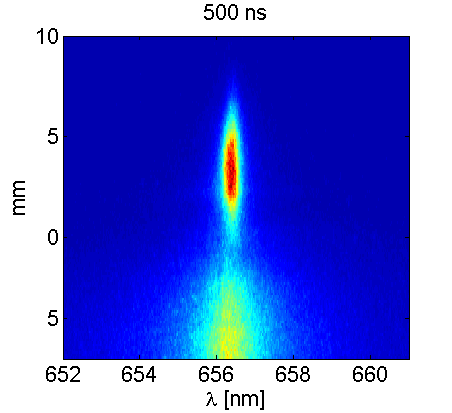}
	\includegraphics[width=0.26 \paperwidth]{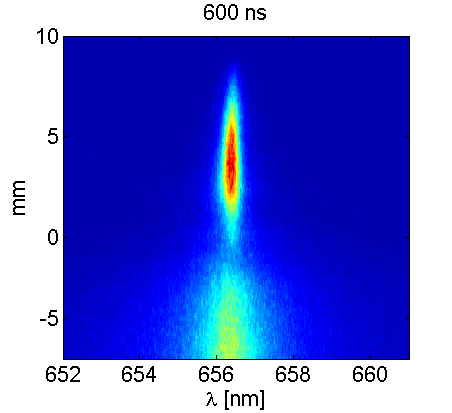}
	
	\caption{Progression of the plasma density outside the capillary near one of the electrodes. The negative values of the vertical axis represent the plasma confined into the capillary. In the pictures it is possible to clearly distinguish the shadow of the electrode from 0 to -1 mm.}
	\label{PlasmaPlumePositive}
	\vspace*{1em}
\end{figure*}
Gas-filled capillaries are widely used in plasma-based acceleration experiments \cite{Spence2000,Pompili2017} where large amplitude plasma waves are excited by the interaction of a laser or a particle beam (referred as \emph{driver} beam) with the plasma. In these devices, the gas is ionized before the interaction with the driver. Moreover, the high current discharge, combined with the use of low atomic number gases like hydrogen, let to reach a high ionization level of the atoms. This let to reduce the ionization losses \cite{Butler2002} which is particularly important for low energy drivers. These capillaries are usually made of hard materials, like alumina, quartz, diamond or sapphire \cite{VanTilborg2015,Butler2002,Gonsalves2016}. Recently polymers have been used \cite{FilippiTBP}. The geometry of the capillary fairly influences the discharge process, while it definitively plays a primary role in the plasma density distribution in the latter instants of the plasma formation.  

Three different stages can be identified in a capillary discharge \cite{Bobrova2001}. In the first stage, the plasma density, the plasma electron temperature and the ionization rate increase as the current increases till the moment of almost full ionization. The current penetrates the plasma very quickly and the plasma is heated and ionized locally due to collisional process. In the second stage, which starts few tens of nanoseconds after the ignition of the discharge, the gas is almost fully ionized and the rising of the electric current allows the increase of the temperature. In the third and last stage of the plasma discharge, plasma reached a quasi-steady-state equilibrium at a given electric current, in which the electron temperature equals the ion temperature (Local Thermal Equilibrium, or LTE plasma). During this last stage, which is the longer and lasts from few hundreds of nanoseconds till some microseconds, the plasma modifies its density distribution also due to the effect of the capillary which confines it. The experimental measurement of the plasma density evolution is always beneficial to understand the plasma-beam interaction process.

In previous measurements we observed that the plasma density profile along the capillary is reproducible shot-by-shot and can be measured along the entire interaction length \cite{FilippiJinst2016}. By analyzing the temporal evolution of the plasma density distribution along the capillary, we observed that the density increases until the complete ionization of the gas, then it starts to decrease being flatter \cite{FilippiIPAC2017}. It is straightforward that it is possible to wait until the plasma reached the expected density profile to let the driver interact with the desired plasma density. For example, in resonant particle-driven plasma acceleration experiment plasma density of the order of $10^{16}$ cm$^{-3}$ is required \cite{Chiadroni2017}. For this experiment it is possible to wait until 1000 ns after the discharge trigger and even more. An example of the plasma density distribution in a 1 cm long, 1 mm diameter capillary for different delays from the discharge trigger is plotted in Fig. \ref{DensityProfile}. Due to the outflow of the plasma, the density inside the capillary (which lies between 0 and 10 mm) decreases while the amount of plasma flown outside increases. The density profile encountered by the laser/particle beam at the latter instants after the ionization is then longer than expected. The interaction path of the driver with the plasma starts much before entering into the capillary and it continues after with a non negligible density which may rise to unwanted (if uncontrolled) effects.

The accurate characterization of these ramps is then mandatory to understand the correct beam-plasma interaction. 

\section{Plasma outflow}
Plasma ramps are formed by the plasma which flows out the capillary after the ionization. Many different processes act on this phenomenon, such as the ion thermal motion and the shock waves of the plasma ionization \cite{Bobrova2001,Biagioni2016}. Ion thermal motion strongly depends on the plasma temperature, since the ion acoustic velocity $c_s$ depends on this parameter ($c_s=\sqrt{T_e/m_i}$) \cite{Bobrova2001}, where $T_e$ is the plasma electron temperature in eV and $m_i$ is the ion mass. Plasma electron temperature is equal to the ion temperature due to the LTE of the plasma. The formation of the shock waves, on the other hand, is much more complicated and it depends on the thermodynamic properties of the ionized gas and the geometry of the capillary. However, the shock wave, due to its nature, has a very short duration and after few hundreds of nanoseconds it does no more affect the plasma flow.

In the following we will introduce some evidences of the plasma outflow and we will discuss about their expansion velocity and plasma density after some hundreds of nanoseconds from the discharge trigger, when only the ion thermal motion acts on plasma outflow.

\subsection{Ramp growth velocity}

The growth of the plasma outside the capillary is evident in the raw spectroscopic images represented in Fig. \ref{PlasmaPlumePositive}. These pictures have been obtained by collecting the light self-emitted by the ionized hydrogen with an imaging system, then imaging it onto a spectrometer with a 2400 g/mm grating which spectrally disperse it \cite{FilippiJinst2016}. A gated intensified CCD camera is triggered in order to acquire the images with 100 ns of gate time by varying its delay respect to the discharge trigger. For these pictures we decide to observe the emission of the Balmer alpha line (656.3 nm) because it has a strong signal in the visible range and this makes it best suited for the analysis of the spatial evolution of the pure hydrogen plasma. In the first image, after 400 ns from the discharge trigger, the plasma outside the capillary is more than 3 mm long and it continuously grows in the following instants represented in the other two pictures. At these time delays, only the ion thermal motion acts on the plasma. The different length of the ramps let to roughly measure the plasma expansion velocity around 13800 m/s, which corresponds to a ion thermal motion of $1.9$ eV. This value is comparable with our simulation \cite{Anania2016} and corresponds to an almost complete ionization of the hydrogen ionized by the discharge.

\subsection{Ramp plasma density}
\begin{figure}
	\begin{center}
		\includegraphics[width=1 \columnwidth, trim=0 0 0 0]{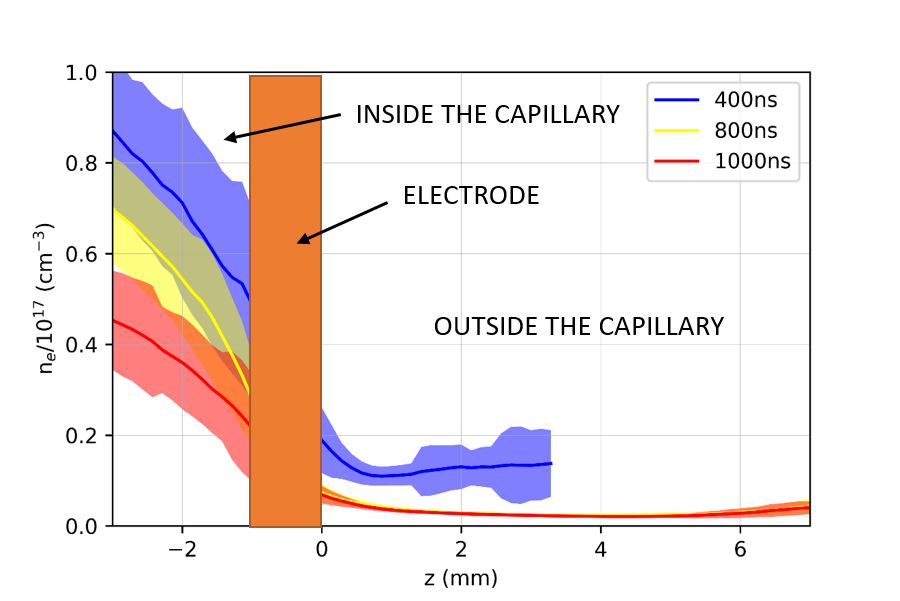}
		\caption{Outflow of the plasma density from the entrance of the capillary (with standard deviation obtained over 200 shots). The lines show the density measured at different delays from the discharge trigger, respectively 400, 800 and 1000 ns.}
		\label{PlasmaOutflow}
	\end{center}
\end{figure}
The plasma density of the ramps has been analyzed with the Stark broadening technique\cite{Gigosos2003}. By using the same optical system used for acquiring the Balmer alpha line it has been possible to analyze the broadening of the Balmer beta (486.1 nm). By measuring its broadening, from the full width at half maximum (FWHM) of this line we retrieved the local plasma density located around the emitter. Balmer beta line, even if weaker than the Balmer alpha, allows for more precise measurements since it is almost insensitive to the plasma temperature until the value of 10 eV \cite{Gigosos2003}. From the measurements exposed above and from the simulations we observed that the plasma temperature is well below this limit. With this system, we reached a spatial resolution of almost 133 $\mu$m and temporal resolution of 100 ns. 
The evolution inside and outside the capillary so measured is plotted in Fig. \ref{PlasmaOutflow}. The plasma density has been acquired for different discharge delays by properly delaying the camera trigger. The shot-to-shot standard deviation error represented by the error bars are obtained by analyzing 200 shots. In the plot, we decided to plot the same delays used to acquire the data in Fig. \ref{DensityProfile}. The length of the plasma ramp increases with the delay, but the density decreases until $3 \times 10^{15}$ cm$^{-3}$ and remains almost constant outside the capillary, probably due to fluid effects caused by the shape of the electrode. 


\section{Conclusions}
We have studied the evolution of the plasma density inside and outside an hydrogen-filled capillary discharge for different time delays from the beginning of the discharge. 

In particular, we have measured the velocity of the plasma expansion by observing the length of the self-emitted light of the Balmer alpha line imaged into an imaging spectrometer. We have also observed the electron plasma density inside and outside the capillary at different delays respect to the discharge trigger with the analysis of the Stark broadening of the Balmer beta line. One thousand nanoseconds after the discharge the plasma ramp has a nearly constant density of the order of $10^{15}$ cm$^{-3}$ extending for more than $8$ mm outside the capillary, more than doubling the effective length of the plasma encountered by the beam. The study of the plasma ramps allows a better understanding of the beam-plasma interaction in future plasma-based experiments. The formation of the ramps, which cannot be neglected in the beam-plasma interaction, must be taken into account in the future projects of plasma targets for plasma accelerators. 

Further analysis will deeper investigate the evolution of the ramps and the mechanism which lies beyond their formation in order to properly control them, preventing from beam quality degradation. 

\section{Acknowledgment}
This project has received fundings from the European Union's Horizon 2020 research and innovation programme under grant agreement No. 653782.

\newpage

\bibliography{mybibfile}

\bibliographystyle{model1-num-names}

\end{document}